\documentclass[prb, twocolumn, superscriptaddress, showpacs]{revtex4-1}
\usepackage{graphicx}
\usepackage{dcolumn}
\usepackage{bm}
\usepackage{amsmath}

\begin{document}
\title{Thermally driven ballistic rectifier}
\author{J.~Matthews}
\affiliation{Physics Department and Materials Science Institute, University of Oregon, Eugene, Oregon 97403-1274}
\author{D.~S\'{a}nchez}
\affiliation{Instituto de F\'{\i}sica Interdisciplinar y Sistemas Complejos IFISC (CSIC-UIB), E-07122 Palma de Mallorca, Spain}
\author{M.~Larsson}
\affiliation{Solid State Physics and The Nanometer Structure Consortium (nmC@LU), Lund University, Box 118, S-221 00, Lund, Sweden}
\author{H.~Linke}
\affiliation{Physics Department and Materials Science Institute, University of Oregon, Eugene, Oregon 97403-1274}
\affiliation{Solid State Physics and The Nanometer Structure Consortium (nmC@LU), Lund University, Box 118, S-221 00, Lund, Sweden}
\begin{abstract}
The response of electric devices to an applied thermal gradient has, so far, been studied almost exclusively in two-terminal devices. Here we present measurements of the response to a thermal bias of a four-terminal, quasi-ballistic junction with a central scattering site. We find a novel transverse thermovoltage measured across isothermal contacts. Using a multi-terminal scattering model extended to the weakly non-linear voltage regime, we show that the device's response to a thermal bias can be predicted from its nonlinear response to an electric bias. Our approach forms a foundation for the discovery and understanding of advanced, nonlocal, thermoelectric phenomena that in the future may lead to novel thermoelectric device concepts.
\end{abstract}


\maketitle
Nanoelectronic devices with dimensions smaller than the characteristic scattering lengths for electrons are easily driven into the nonlinear regime, and show electron dynamics that depend on device symmetry. For instance, clear rectification effects have been
observed in asymmetric microjunctions \cite{Song1998,Dehaan2004,Hackens2006}, quantum dots \cite{Linke1998} and
quantum point contact systems \cite{Linke1999,Shorubalko2001}. These systems exhibit 
non-equilibrium effects that occur in the {\em nonlinear} regime of ballistic transport.

At the same time, there has been considerable progress in the use of
mesoscopic conductors for thermoelectric devices such as nanoscale coolers
and ultrasensitive thermometers \cite{Giazotto2006}.
At a more fundamental level, studies of heat flow in phase-coherent systems
allow one to investigate basic properties of quantum transport when electric
and thermal gradients act jointly on the same system \cite{Dubi2011}.
Specifically, thermoelectric effects are observed when
a voltage drop builds up in response to a temperature difference
\cite{Molenkamp1990,Godijn1999}. Thus far, most of the literature
has investigated thermopowers in two-terminal arrangements
(for exceptions see Refs. \cite{Molenkamp1990,Pogosov2002}).

Additionally, for the past four decades there has been sporadic research on so-called anisotropic thermoelectric materials. These are materials that exhibit off-diagonal elements in their electrical conductivity, thermal conductivity, or thermopower. Typically, anisotropic thermoelectrics comprise of either stacked multilayers of alternating materials \cite{Babin1974,Kyarad2004}, or anisotropic crystalline materials \cite{Zahner1997}. Such systems, however, can only change the anisotropic properties of a given system by rotating the stacked multilayer or crystal orientation. This effectively limits the range of anisotropic properties that can be studied.

\begin{figure}[b]
\begin{center}
\includegraphics{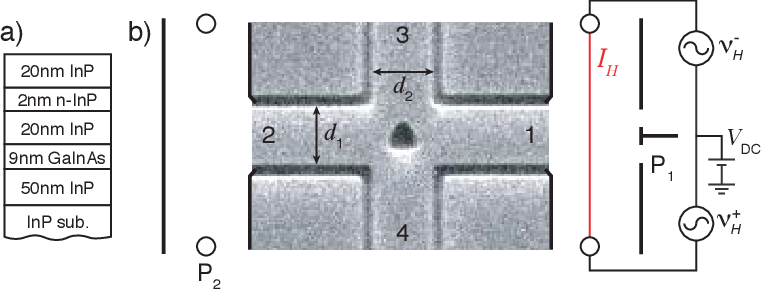}
\caption{(a) Schematic of the wafer structure with a 2DEG located in the GaInAs quantum well. (b) Schematic and SEM image of a four terminal antidot junction similar to the one reported on here. The four circles indicate connections to distant Ohmic contacts, not shown. The circuit diagram on the right shows the heating circuit used to heat the side channels; the configuration shown here is used to heat the right side channel. The lateral and transverse voltage drops are given by $V_{12}$ and $V_{34}$, respectively. The actual four-terminal junction measured here had terminal widths of $d_1\approx500$ nm and $d_2\approx450$ nm, and an antidot with a base and height of about 550 nm and 380 nm, respectively. The side heating channels are 5 $\mu$m wide, 30 $\mu$m long, and $\approx$ 4 $\mu$m from the scatterer. Note that the side channels are not to scale.}
\label{fig:AntidotSchematic}
\end{center}
\end{figure}

Here, we address transverse thermoelectric effects in a quasi-ballistic device fabricated in a two-dimensional electron gas (2DEG), Fig.~\ref{fig:AntidotSchematic}(a), which offers the great advantage that the scattering properties of the system can be controlled during fabrication. Specifically, we use a four-terminal microjunction with a central, asymmetric antidot to scatter electrons, Fig.~\ref{fig:AntidotSchematic}(b), and perform thermoelectric measurements at a cryostat temperature of 0.74 K. In a similar device, previous observations have shown that a lateral AC \emph{voltage} leads to a rectified, transverse response \cite{Song1998}. Here, we instead apply a lateral \emph{temperature} difference 
across the junction through terminals 1 and 2, and observe 
a nonlocal thermovoltage between terminals 3 and 4, both of which are unheated. 
We note that similar thermovoltages have been detected previously
in multi-terminal 2DEGs subjected to magnetic fields normal to the system
(the Nernst effect \cite{Pogosov2002,Maximov2004,Goswami2011}). In contrast to these works, the effect reported here is purely thermoelectric in nature and only due
to the symmetry breaking property of the central scatterer.
Furthermore, we establish a theoretical model based
on the multi-terminal scattering approach \cite{Buttiker1986,PNButcher1990}
extended to the weakly nonlinear regime (up to quadratic order in voltage
and linear order in temperature shifts). We then show that this model can predict the transverse thermoelectric response to a lateral thermal bias from nonlinear voltage measurements similar to those reported in Ref.~\cite{Song1998}.

Our device consists of an InP/Ga$_{0.23}$In$_{0.77}$As 2DEG wafer, see Fig.~\ref{fig:AntidotSchematic}(a), patterned using the same techniques as in Ref.~\cite{Larsson2008}. The antidot junction itself and its dimensions are shown in Fig.~\ref{fig:AntidotSchematic}(b). Hall and Shubnikov de Haas measurements give a carrier concentration and mobility of $2.93\times10^{12}$ cm$^{-2}$ and $1.16\times10^5$ cm$^2$/Vs respectively. The resulting mean free path, 3.3 $\mu$m, exceeds the characteristic length of the antidot region, $0.5$ $\mu$m. Therefore, electrons within the antidot junction are in the ballistic regime.

\begin{figure}[t]
\begin{center}
\includegraphics{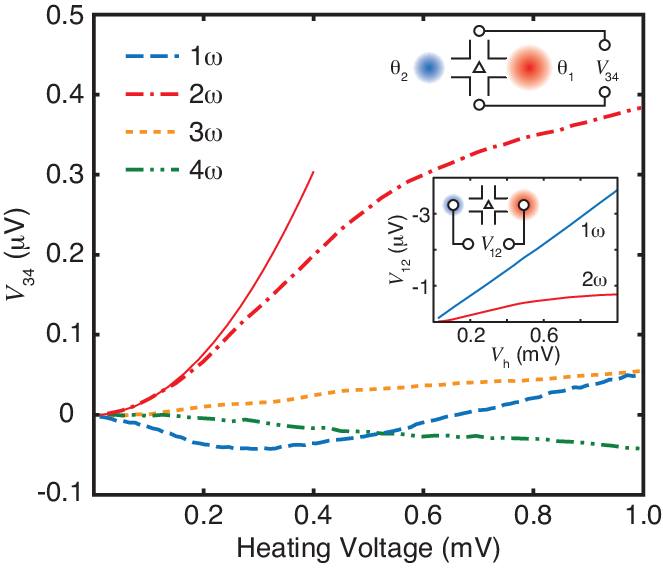}
\caption{(Main figure) First four harmonic transverse voltage responses to a temperature increase in terminal 1. The solid red line is a parabolic fit to the 2$\omega$ curve at low voltages, and is given by $V_{34}^{(2\omega)} = (1.9\text{V}^{-1})V_h^2$. (Top inset) Schematic of the measurement configuration. (Bottom inset) Measured lateral voltage drop across the junction during heating.}
\label{fig:ThermalResultsHarmonics2}
\end{center}
\end{figure}

For our thermoelectric measurements, we are primarily interested in the transverse voltage response, $V_{34}=V_3-V_4$, due to a change in the side terminal temperature(s), of either $\theta_1$, $\theta_2$, or both. Here $V_i$ and $\theta_i$ are the shifts of the electrochemical potential and temperature in the $i^{th}$ terminal away from the common electrochemical potential, $\mu/e$, and background cryostat temperature, $\theta$. To increase either $\theta_1$ or $\theta_2$, we apply two 37 Hz heating currents, which have a relative phase difference of 180$^{\circ}$, to the channel to be heated. Figure~\ref{fig:AntidotSchematic}(b) shows the circuit diagram for heating the right side channel. Using $P_1$ as a voltage probe, the relative amplitudes of $\nu_H^+$ and $\nu_H^-$ are tuned so that the resulting 1$\omega$ voltage at terminal 1 was minimized. Any remaining DC offset could have been minimized as well using a DC source to shift both heating voltages; however, no offset was needed. This technique allows us to apply a thermal bias without an electrical bias.

\begin{figure}[t]
\begin{center}
\includegraphics{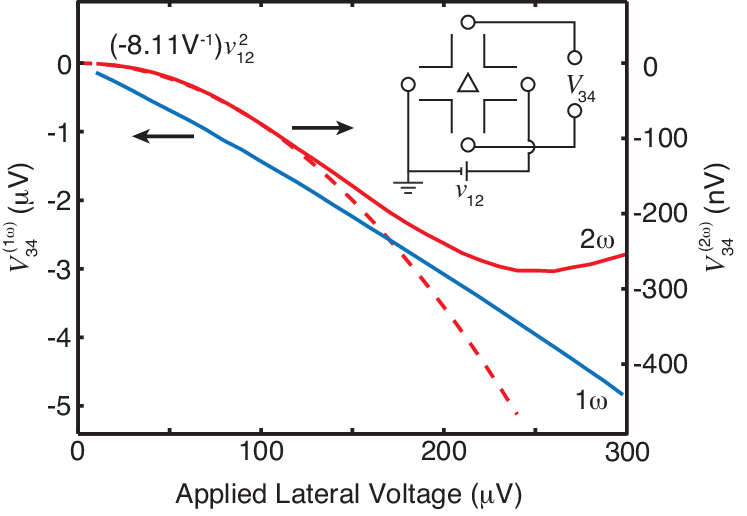}
\caption{(Main figure) 1$^\mathrm{st}$ and 2$^\mathrm{nd}$ harmonic transverse responses, $V_{34}$, to a laterally applied 37 Hz voltage $\nu_{12}$. Also shown is a parabolic fit to the 2$\omega$ curve at low voltages. (inset) Circuit schematic for the applied voltage measurement.}
\label{fig:VoltageResults}
\end{center}
\end{figure}

In Fig.~\ref{fig:ThermalResultsHarmonics2} we show results from an experiment where a temperature shift $\theta_1$  was applied to terminal 1, and the first four harmonics, with respect to the heating voltage, of $V_{34}$ were measured. Since the thermal gradient is generated from Joule heat, which is primarily proportional to the square of the applied heating voltage, the dominant 2$\omega$ response implies that the transverse response is proportional to the applied temperature gradient, $V_{34}\propto{}V_H^2\propto{}\theta_1$. We also measured the first and second harmonics of the lateral voltage drop $V_{12}$ using the probes labeled $P_1$ and $P_2$ (see inset to  Fig.~\ref{fig:ThermalResultsHarmonics2}). The lateral voltage by itself, $V_{12} < 4$ $\mu$V, is far too small to cause the observed transverse voltage, as can be seen from the device's measured 1$\omega$ and 2$\omega$ responses to an applied 37 Hz voltage, see Fig.~\ref{fig:VoltageResults}. 

At low voltages, $V_{34}$ in Fig.~\ref{fig:ThermalResultsHarmonics2} is linear in $\theta_1$ and can thus be interpreted as a thermovoltage. Remarkably, however, this thermovoltage is across two unheated contacts that, based on device geometry, can be assumed to be isothermal. We therefore propose that $V_{34}$ is due to the difference between two or more thermovoltages in the junction. Sources of thermovoltages that are easily identifiable are the four quantum point contacts (QPC) defined by the antidot and the surrounding channel walls. In the general case of zero symmetry, each QPC thermovoltage would be unique, resulting in all four terminals being at different potentials. We will return to this interpretation below. 

\begin{figure}[t]
\begin{center}
\includegraphics{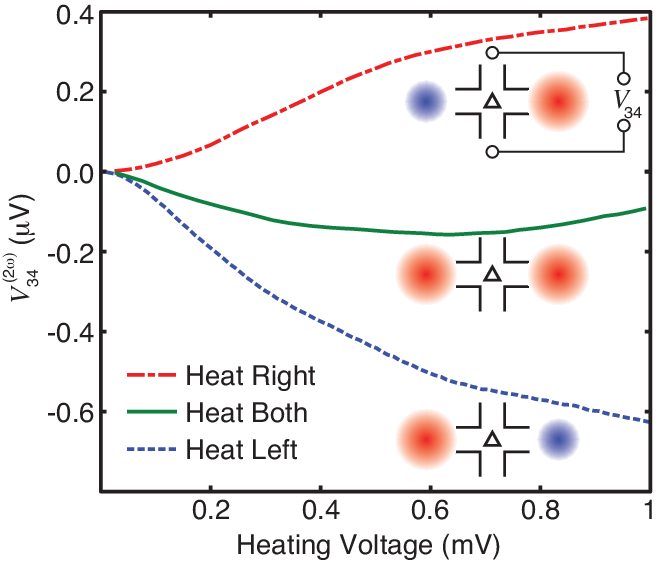}
\caption{$2\omega$ transverse response due to heating both side channels simultaneously and individually. The vertical dotted lines indicate the approximate onsets of nonlinear regimes.}
\label{fig:ThermalResultsHeatBoth}
\end{center}
\end{figure}

Given the intended left-right geometric symmetry of the junction one may intuitively expect that $V_{34}(\theta_1,\theta_2)=V_{34}(\theta_2,\theta_1)$. When the lateral temperature differential is reversed, however, measurements show a sign reversal in $V_{34}$ (Fig.~\ref{fig:ThermalResultsHeatBoth}). Furthermore, when both terminals are heated simultaneously, the response tends toward zero. We will return to these unexpected observations shortly.

We now aim to explain our results using the multi-terminal thermoelectric scattering approach
\cite{Buttiker1986,PNButcher1990} extended to the weakly nonlinear transport regime \cite{Christen1996}. We consider a four-terminal junction with a central scatterer, as in Fig.~\ref{fig:AntidotSchematic}(b) and Ref.~\cite{Song1998}, but without assuming any particular model for the scattering matrix.
Based on the linear temperature and quadratic voltage responses shown in the experiment
(see Figs.~\ref{fig:ThermalResultsHarmonics2} and~\ref{fig:VoltageResults}),
we express the current $I_\alpha$ through lead $\alpha$ up to linear order in temperature
and quadratic order in voltage differences, $V_{\alpha\beta}=V_{\alpha}-V_{\beta}$:
\begin{align} \label{eqn:LBTheory}
I_{\alpha}=\sum_{\beta\neq\alpha}[&G_{\alpha\beta}(V_{\alpha\beta})
+F_{\alpha\beta}(V_{\alpha\beta})^2+L_{\alpha\beta}(\theta_{\alpha}-\theta_{\beta})]\,.
\end{align}
Importantly, Eq.~\eqref{eqn:LBTheory} is manifestly gauge invariant, a property
recently emphasized for mesoscopic rectifiers \cite{Buttiker90}.

In the low temperature limit, the linear transport coefficients read,
\begin{align}
G_{\alpha\beta}&=\frac{2e^2}{h}t_{\alpha\beta}(\epsilon_F)\\
L_{\alpha\beta}&=\frac{2\pi^2ek_B^2\theta}{3h}t'_{\alpha\beta}(\epsilon_F).
\end{align}
This approximation is valid provided the energy variation of the transmission probability,
$t_{\alpha\beta}$, between leads $\alpha$ and $\beta$
is weak and the Fermi energy, $\epsilon_F$, is larger than the background temperature, $\theta$.
This amounts to neglecting band edge effects. Hence, the transmission
and its energy derivative, $t'_{\alpha\beta}$, are evaluated at $\epsilon_F$.

The leading order rectification term $F_{\alpha\beta}$
is a complicated function of $t'_{\alpha\beta}$ and of the transmission
derivative $\partial t_{\alpha\beta}(E,\{V_i\})/\partial V_\alpha|_{\rm eq}$
evaluated at equilibrium \cite{Christen1996}.
Knowledge of the latter requires a self-consistent determination
of the screening potential, $U(\{V_i\})$, inside the conductor. However,
within a qualitative discussion we can neglect this term
since it is proportional to the characteristic potential,
$u_\alpha=(\partial U/\partial V_\alpha)|_{\rm eq}$ \cite{Christen1996},
whose strength can be estimated within mean-field theory from $u\sim C_\mu/C$, where
$C_\mu^{-1}=(e^2 D)^{-1}+C^{-1}$, $C$ is a capacitance coefficient that measures interactions,
and $D$ denotes the density of states associated with the scattering states.
In widely open systems one has $C\gg e^2D$ (noninteracting limit),
and the rectification term becomes
\begin{equation}
F_{\alpha\beta}\simeq-\frac{e^3}{h}t'_{\alpha\beta}(\epsilon_F).
\end{equation}

To find $V_{34}$ as a function of the lateral voltages ($V_1$ and $V_2$)
and temperatures ($\theta_1$ and $\theta_2$) we recall that terminals 3 and 4
act as voltage probes. We then solve the two conditions $I_3=I_4=0$
with the aid of Eq.~\eqref{eqn:LBTheory} in the cold isothermal case ($\theta_3=\theta_4=0$), that is, we assume that the channel thermal conductances are much greater than the thermal conductances across the antidot region.
For definiteness, we neglect the products $V_{i}V_{(3,4)}$
since in our experiment,
$V_{34}$ is much smaller than $V_{12}$. Then,
\begin{equation}\label{eqn:Theory}
V_{34}=A(V_1-V_2)+\sum_{j=1,2} B_j (g_{\theta} \theta_j - g_e V_j^2)\,,
\end{equation}
where we have defined,
\begin{align}
A &= \frac{t_{31}t_{42}-t_{41}t_{32}}{D}\,,\label{eqn:A} \\
B_j &= \frac{(t_{41}+t_{42})t'_{3j}-(t_{31}+t_{32})t'_{4j}}{D}\,,\label{eqn:B}
\end{align}
with $D=(t_{31}+t_{32})(t_{41}+t_{42})+t_{34}(t_{41}+t_{42})+t_{43}(t_{31}+t_{32})$.
Here $g_e= e/2$ and $g_{\theta}=\pi^2k_B^2\theta/3e$. 

Interestingly, Eq.~\ref{eqn:Theory} predicts that the coefficients $B_j$ that describe the non-linear voltage response also describe the transverse response to a laterally applied temperature differential. We can utilize this relationship to extract the expected thermopower from the electrical measurements in Fig.~\ref{fig:VoltageResults}, and the electron temperature rise from the thermal bias measurements in Fig.~\ref{fig:ThermalResultsHarmonics2}.

From the parabolic fit in Fig.~\ref{fig:VoltageResults}, and Eq.~\eqref{eqn:Theory}, we determine $B_1=2.02\times10^{20}$ J$^{-1}$. Inserting this value back into Eq.~\eqref{eqn:Theory} yields a transverse thermopower of $S_{34,12}=B_1g_\theta=594$ nV/K. We will address the magnitude of this thermopower in a moment.

With $S_{34,12}$ in hand, we can then estimate $\theta_1$ as a function of $V_h$. Using the parabolic fit from Fig.~\ref{fig:ThermalResultsHarmonics2}, we find $\theta_1\approx (3.2\times 10^6 \text{ K/V}^2) V_h^2$, which is valid for $V_h<0.2 \text{ mV}$. At the upper limit of the fit, $V_h=0.2$ mV, this yields $\theta_1\approx130$ mK. These temperatures agree well with a simple diffusive-heat finite-element simulation of the electron temperature in the side channel: $\theta_1^{\text{sim.}} \approx (2.5\times 10^6 \text{ K/V}^2) V_h^2$ for $V_h<0.2$ mV.

To offer a more conceptual description of our observations, we now interpret the effective thermopower of the junction as the difference of QPC thermopowers between neighboring terminals. For the configuration where heat is applied to the right channel, we interpret the effective thermopower as the difference between the top-right and bottom-right QPC thermopowers. Correspondingly, we interpret the thermopower for heating of the left channel as due to the left-hand pair of QPCs in the junction. To check whether this interpretation is reasonable, we compare the magnitude of $S_{34,12}$ to the difference between two ideal QPC thermopower peaks:
\begin{equation} \label{eqn:S_QPC}
\Delta{}S^{QPC}=-\frac{k_B\ln(2)}{e}\left (\frac{1}{N_{T}-0.5}-\frac{1}{N_{B}-0.5} \right ),
\end{equation}
where $N_{(T,B)} \geq1$ represent the integer number of conductance modes in the respective top and bottom QPCs. Using the measured two-terminal conductance between terminals 1 and 2, $G_{12}^{meas.}=8.73\times10^{-4}$$\Omega^{-1}$, we estimate the number of modes to be around 11. For mode numbers between 10 and 13, Eq.~\eqref{eqn:S_QPC} predicts two-QPC thermopowers with a magnitude between $0.4$ and $1.5$ $\mu$V/K, which fits well with $S_{34,12}$. Here it is important to note that the true two-QPC thermopower is a function that strongly oscillates between positive and negative values \cite{Molenkamp1990}, and Eq.~\eqref{eqn:S_QPC} only predicts the range of reasonable values. However, the oscillatory nature of this thermopower offers a simple explanation for the sign change in the thermopower observed in Fig.~\ref{fig:ThermalResultsHeatBoth}: because the sign of the thermopower depends sensitively on the number of occupied modes in each QPC, even a very small left-right asymmetry in the width (occupation number) of the QPCs in our ballistic junction can lead to a sign change.

Our conceptual two-QPC model also offers an explanation for the nonlinear changes in slope of the thermopower as a function of heating power (see vertical, dashed lines in Fig.~\ref{fig:ThermalResultsHeatBoth}), which have been seen in similar two-QPC systems \cite{Molenkamp1990}. As the temperature drop across a QPC increases, thermal broadening about the Fermi energy accesses neighboring thermopower peaks and increases the QPC's thermopower. The downward change in slope in the heat-right curve can then be interpreted as an increase in the number of accessed thermopower peaks in the top right QPC as a function of $\theta_1$. Similarly for the heat-left configuration, the two nonlinear points indicate increases in the number of accessed thermopower peaks in the bottom left QPC.

It is important to note that there are in fact four interacting QPCs in the junction, a point the two-QPC model partially neglects. To properly interpret the observed sign change in Fig.~\ref{fig:ThermalResultsHeatBoth} then, we refer to \eqref{eqn:Theory}, which simultaneously accounts for all four QPCs. Focusing only on the response to temperature changes, the $2\omega$ response can be decomposed into components that are symmetric, $S_{34}^S$, and antisymmetric, $S_{34}^A$, under the transformation $(\theta_1,\theta_2) \to (\theta_2,\theta_1)$:
\begin{equation}
V_{34}=S_{34}^S\frac{\theta_1+\theta_2}{2} + S_{34}^A\frac{\theta_1-\theta_2}{2}\,,
\end{equation}
where $S_{34}^S=B_1+B_2$ and $S_{34}^A=B_1-B_2$. When both terminals are heated simultaneously and equally, the main contribution stems from the symmetric component, which is generally nonzero. When $\theta_1$ and $\theta_2$ are not equal and exchanged, the symmetric component remains unchanged while the antisymmetric component reverses sign. Note that $S_{34}^A$ exists only for scatterers showing some kind of asymmetry in the transmission derivative coefficients ($t_{31}'\neq t_{32}'$ or $t_{41}'\neq t_{42}'$). Since we observe a complete change in sign, our device must have a strong asymmetric component.

We have shown that a four-terminal junction with a symmetry-breaking scatterer can be used to generate a thermovoltage between two cold isothermal contacts. We have put forward a simple noninteracting model for multi-terminal thermoelectric transport that accounts for most observed effects. 
Given the great control over device geometry and symmetry afforded by nanofabricated devices, our approach can be used to explore advanced thermoelectric functionalities --- such as electronic thermal rectifiers or more efficient thermoelectric energy converters --- due to the opportunity to separate heat and charge flow in multi-terminal devices.

We acknowledge Hongqi Xu of Lund University for fruitful discussions, and financial support from NSF IGERT grant No. DGE-0549503, the National Science Foundation Grant No. DGE-0742540, ARO Grant No. W911NF0720083, Energimyndigheten Grant No. 32920Ð1, MICINN Grant No. FIS2011-23526, nmC@LU, and the Foundation for Strategic Research (SSF). Effort sponsored by the Air Force Office of Scientific Research, Air Force Material Command, USAS, under grant number FA8655-11-1-3037. The U.S. Government is authorized to reproduce and distribute reprints for Governmental purposes notwithstanding any copyright notation thereon.

\end{document}